\documentstyle[12pt]{article}
\begin{document}
\thispagestyle{empty}
\begin{flushright}
CTP-TAMU-11/94\\
February 1994
\end{flushright}
\begin{center}
\begin{LARGE}
{\bf A Vision for High Energy Physics}\\[.2in]
\end{LARGE}
\begin{large}
Teruki Kamon\\
Jorge Lopez\\
Peter McIntyre\\
James White\\[.2in]
Texas A\&M University\\
Department of Physics\\
College Station, TX 77832--4242\\
\end{large}
\end{center}
\begin{abstract}
{\small\baselineskip=10pt
Following the termination of the Superconducting Super Collider, there is an
urgent need to develop a strategic plan for the future of high energy physics
and an accompanying vision to guide the priorities of the U.S. program.
This document proposes such a strategic plan and presents a singular
opportunity for the U.S. program. The existing hadron collider at Fermilab
could be upgraded to create a major discovery potential for supersymmetry,
one of the most profound concepts in the world of elementary particles.
Using a single ring of SSC magnets in the existing tunnel, the recently
improved understanding of SUSY phenomenology, and the upgraded detectors
in place at the Tevatron, the DiTevatron could be doing physics within five
years and reach most of the range of parameters permitted in SUSY models.
We propose that it be funded as a worthy component of the SSC termination,
bringing to fruition both the technology and the science of the SSC at very
modest cost.
}
\end{abstract}
\vspace{.4in}

        With the advent of the Superconducting Super Collider, the U.S. high
energy physics community embarked on a quest which would have taken a decade to
begin
operation, and aspired to leap more than an order of magnitude in all measures
of performance beyond anything that exists today.  Its charter mission was to
search for the
Higgs boson - the carrier of a scalar field which is thought to be the origin
of mass for the
gauge bosons of the electroweak interactions. With the untimely termination of
the
project, our community faces one of the biggest turning points in its history.
There is
good news and there is bad news.  To the good, the search for the elusive top
quark is
reaching a climax, with the focus of attention at Fermilab.  To the bad, the
aggregate U.S.
funding of high energy research, $\sim$\$650 million/year, is likely to be cut
by 4\% next year
and prospects are little better in future years.

{\bf Can we develop a realistic strategic plan which has the potential for
milestone discovery within the next $\sim$6 years?}  If our field of science is
to prosper,
indeed if it is even to survive, we must develop a vision of the future which
can earn the
public support which we have come to expect.  At the present budget level, the
U.S.
would fund nearly \$4 billion to high energy research in that time.
Ironically,
that is the
original cost estimate for the SSC itself.  The American people, their
government, and our
colleagues in other fields of science expect that we develop a strategic plan
which has
reasonable prospect of yielding new discoveries about nature which are
commensurate
with this level of support.  If we fail to develop such a plan, or if we fail
to communicate it
effectively, we will likely receive less support in each succeeding year.

\noindent{\bf Historical context.}    In order to set a context for milestone
discovery,
we can examine
the history of our field.  Table I shows a chronology of such discoveries in
elementary
particle physics.  Each of these discoveries has transformed the way we
understand the
fundamental particles and interactions of nature.  The field began with the
discoveries of
natural radioactivity in 1896 and the electron in 1897, Rutherford scattering
from the
nucleus in 1911, and the neutron in 1932.  After World War II a steady stream
of
milestone discoveries, averaging one every five years, has brought the field
to its current
understanding of the Standard Model.  Most of these milestones have been
recognized by a Nobel Prize.

        The latest such milestone may be just around the corner: the top quark.
The top
quark is the ``missing link" in the bestiary of fermions in the Standard Model.
{}From the
measurement of the $Z^0$ boson decay width at LEP (one of the latest milestone
discoveries), we
believe the top quark to be the last Standard Model fermion to be discovered.
The search
for top is the primary focus of effort in the CDF and D0 experiments at
Fermilab, and both
experiments will be announcing new results this Spring.  The efforts of both
collaboration are distinctly international in character.  CDF includes groups
from Canada, Italy, Japan, and the US; D0 is a collaboration including  Brazil,
Colombia, France, India, Mexico, Russia,  and the US.

        As we look to the future, it is of course pointless to resort to the
oxymoron of
predicting discoveries.  We will however be judged in the context of the
revolutionary
advances we have made in the past forty years, and of the enormous cost of our
future
research.  We must therefore look with clear vision at the potential for
milestone discovery
which attends the several elements of the world high energy research program,
and target
the U.S. program to maximize that potential, recognizing the constraints of
funding at home and the legitimate ambitions of our colleagues in Europe and
Japan.

\noindent {\bf The potential for future milestone discoveries.}  The task of
assessing  the potential for milestone discoveries is guided by our concepts
for the major unsolved puzzles of the
time, and the concepts which might solve them or provide fundamental insights.
Three of
the most important examples of such concepts today are the Higgs field, which
would
explain the mystery of mass in gauge theories; the origin of CP violation, a
puzzling
dimension of the gauge fields which may explain the imbalance of matter and
antimatter in
cosmology; and supersymmetry, a new fundamental symmetry which treats fermions
and
bosons in a unified way, and actually relates the internal symmetries to
space-time itself.
Our vision of discovery potential for each of these concepts is however
clouded by our
uncertainty about the masses of the carriers of the new fields and their
couplings to the particles we know.

        The pursuit of the Standard Model Higgs boson is the central goal of
the most
ambitious projects that each sector of the world high energy physics community
can
muster.  The parameters of energy ($40\,{\rm TeV}$) and luminosity
($10^{33}\,{\rm cm}^{-2}\,{\rm s}^{-1}$) for
the SSC
were chosen to cover 100\% of the parameter space within which our concept of
the Higgs
could be valid.  The supersymmetric Higgs is the goal of LEP II, the energy
upgrade of
CERN's $e^+e^-$ collider.  The LHC collider being planned at CERN similarly
targets the
Higgs: its parameters ($14\,{\rm TeV}$ energy and
$2.5\times10^{34}\,{\rm cm}^{-2}\,{\rm s}^{-1}$)
are bounded by the
existing LEP tunnel, but can provide $\sim$70\% reach over the Higgs'
parameter space.

        For CP violation, it will be necessary to precisely measure three
amplitudes (the
so-called unitarity triangle) in the Kobayashi-Maskawa matrix which describes
the weak
couplings between the quark families.  The B factories now being designed at
KEK and
SLAC could measure two of the three amplitudes.  With the planned Injector
upgrade, the
Tevatron experiments could contribute the third amplitude ($b\to s$) through
their studies of $B_s$
decays.  Such a coordinated effort has the long-term potential to explore
CP violation
where it may be strongest.  There will also soon be an opportunity to extend
the study of
CP violation in the strange quark sector at the $\varphi$ factory which is
being built
at INFN Frascati.

        Supersymmetry (SUSY) was originally proposed two decades ago in a
rather abstract
context.  It would couple the two classes of fundamental particles, the
fermions and bosons, whose utterly distinct characters have been one of the
deepest
puzzles of physics.
Supersymmetry has shown up again and again in many places where it wasn't
expected: the solution to the hierarchy problem and the unification of the
gauge couplings in grand unified theories, the radiative breaking of
electroweak symmetry,
the unification
with gravity (supergravity), and in superstrings.  Moreover, supersymmetric
models
contain a natural candidate for the dark matter puzzle in astrophysics: the
lightest
supersymmetric particle (LSP).  Its potential discovery and its impact on our
understanding of nature would be as fundamental as the gauge theories
themselves.  This
significance is in contrast with the many extensions of the Standard Model
that have been
suggested, which do no more than extend its gauge group and matter content.

\noindent{\bf The opportunity for supersymmetry.}  During the past year a
remarkable
opportunity to
seek supersymmetry has arisen, even while the top is being chased, the SSC
was being
terminated, and the European and Japanese high energy communities are preparing
their
plans for massive colliders to chase the Higgs.  Recent developments in
accelerator
technology, phenomenology, and detector performance have converged to
suggest a singular opportunity to extend the search for supersymmetry to cover
$\sim$80\% of
its parameter space.  The opportunity is one which could be mounted at
Fermilab, using a
modest upgrade of its present facilities, and commencing physics operation
within 5 years.

        The accelerator technology is the SSC magnet.\footnote{ P. Wanderer,
``Status of Superconducting Magnet Development," 1993 Particle
Accelerator Conference, Washington (1993),  p.2726.}
A single ring of SSC magnets,
installed in the Tevatron tunnel, could produce $\bar{p}p$ colliding beams
with $4\,{\rm TeV}$ collision
energy and $2\times10^{32}\,{\rm cm}^{-2}\,{\rm s}^{-1}$ luminosity - the
DiTevatron or
SUSYTRON.  By
operating a ring of SSC magnets at $3.5^\circ K$ (feasible with the present
Tevatron cryogenic
system), a collision energy of $3.5\,{\rm TeV}$ would be produced.  Upgrade of
the
cryogenic
plant would enable refrigeration to $1.8^\circ K$, and extend operation to
$4\,{\rm TeV}$ collision
energy.  As we will show below, this relatively modest energy upgrade would
extend the
reach of Fermilab for supersymmetry from $\sim$30\% of the parameter space
with the injector-upgraded Tevatron to $\sim$80\% coverage with the DiTevatron.

        The DiTevatron could utilize the Tevatron for injection at $900\,{\rm
GeV}$.
Its magnets
would then require only a 2:1 dynamic range of operation, eliminating many
of the more
challenging problems of magnet performance required for the 10:1 dynamic
range in SSC
or LHC.  The $5\,{\rm cm}$ aperture of the SSC magnets would be fully adequate
in
this case, including provision for the corkscrew trajectory of the two
separated
counter-rotating
beams.  The magnet length would be chosen to match the Tevatron lattice
and sagitta.

        Figure~\ref{Figure1} shows the arrangement of the DiTevatron and the
Tevatron in the
Fermilab tunnel.  The beam height of the collision point would remain the
same - the
detectors would not have to move.  The Tevatron beam would also pass through
the
detectors, a minor inconvenience, but it would carry beam only while the
collider was
being filled with beams before each store.

        The SSC magnets are industrialized, and a magnet factory capable
of mass
producing them stands ready and idle in Louisiana.  The production tooling
and test
facilities could be modified easily for the magnet length appropriate for
DiTevatron.  The
factory could be placed in operation within a year, and could produce the
required $6\,{\rm km}$ of
magnets in a year of production.  This industrialization and its use to
propel U.S. high
energy research into a potential for milestone discovery would be fitting
and achievable
legacies for the brave vision of the Supercollider.  We propose that the
funding for the
DiTevatron be included in the SSC termination; we can thereby transform
the failure to
build the SSC into the successful use of its technology to create an equally
important discovery potential.

        The phenomenology of supersymmetry has benefited from the steady
improvement in the bounds on the top quark mass,  the improving measurements
of a host
of processes which bound the allowable range of parameters for the masses and
couplings
of supersymmetric particles, and the selection of theoretically well-motivated
supersymmetric models.\footnote{ For a recent review, see J.L. Lopez, D.V.
Nanopoulos, and A. Zichichi, ``Status
of the Superworld: From Theory to Experiment," CERN preprint CERN-TH.7136/94.}
Building upon this expertise, we are
conducting a systematic
analysis of the signals which are being used today to seek supersymmetry at CDF
and D0,
and the cross-section $\sigma$ and branching
ratio $B$ for them at the DiTevatron.  Two
processes have been considered: the production of a pair of the super-partners
of the weak
bosons (Chargino $\chi_1^\pm$, Neutralino $\chi_2^0$),
resulting in a characteristic trilepton decay
signature; and the production of a pair of gluinos and squarks
($\tilde g\tilde g$, $\tilde q\tilde q$,
$\tilde g\tilde q$), resulting in an
equally characteristic signature of multiple jets and missing momentum.

        Figure~\ref{Figure2a} shows an example of a trilepton event observed in
CDF,\footnote{CDF Collabollation, ``SUSY Search Using Trilepton Events from $p
\bar{p}$ Collisions at $\sqrt{s}=1.8$ TeV," XVI International Symposium on
Lepton-Photon Interactions, Cornell University, Ithaca, NY, 10-15 August 1993.}
although this particular event has been rejected in the SUSY analysis because
its third lepton escapes into the end plug of the detector.
Figure~\ref{Figure2b} shows another trilepton event, observed in
D0,\footnote{J. T. White (D0 Collaboration), to appear Proc. of 9th Topical
Workshop on Proton-antiproton Collider Physics, Tsukuba, Japan, 18-22 October
1993.} which however passes all the cuts. The following figures show the yield
$\sigma B$ for the trilepton signature for three supergravity models which
bound the range of theoretical framework:
standard SU(5) (Figure\ref{Figure3}), and no-scale string-inspired SU(5)xU(1)
models with moduli (Figure~\ref{Figure4}) and dilaton (Figure~\ref{Figure5})
supersymmetry breaking at the unification scale.  In each
case, the dotted patterns show the extent of the parameter space, while the
dashed line
shows the discovery limit (95\% CL) for the trilepton mode, extrapolating the
actual
experience in this search at CDF to date (present limit indicated by a solid
line), and
assuming a 3-year data run ($5\,fb^{-1}$).  In each case, the Tevatron with
Injector Upgrade
covers $\sim$30\% of the parameter space; the DiTevatron covers virtually the
entire parameter space. By comparison, LEP II should be able to observe
chargino masses only up to 100 GeV.

In order to determine the reach of the DiTevatron for squarks and gluinos,
we studied two typical cases: (i) gluino pair production with significantly
heavier squarks, as expected in the standard SU(5) model, and (ii) all gluino
squark production channels such that $m_{\tilde q}\approx m_{\tilde g}$, as
expected in the SU(5)xU(1) models. Figure~\ref{Figure6} shows the missing
$p_\perp$ distribution  calculated for gluino pairs (case (i)), and the
distribution of events from the dominant background: $Z\to\nu\bar{\nu}$.
Simple selection criteria were applied (missing $p_\perp > 150\,{\rm GeV}$, 4
jets with $p_\perp > 40\,{\rm GeV}$) yielding a statistical significance
$S/\sqrt{B} = 52(13)$ for a $400(500)\,{\rm GeV}$ gluino mass.
The above signal is only for the gluino; if the squark has a comparable mass
(case (ii)), we would see $\sim7$ times more signal. An estimate of the
reach for these two cases is shown in Figure~\ref{Figure7}, in terms of
the statistical significance $S/\sqrt{B}$. The figure shows that the reach
exceeds 500 GeV for gluino pairs (case (i)) and $\sim700$ GeV for all
gluino-squark combinations (case (ii)). With optimized selection cuts, the
reach could be larger.  This covers all but a small tail of parameter space in
the standard SU(5) model, and $\sim80\%$ of the parameter space in
the SU(5)xU(1) models. The analogous reach for the Tevatron with Injector
Upgrade is $\sim250\,{\rm GeV}$.

        The third development is the coming of age of the detectors
CDF\footnote{ CDF Collaboration, ``{\em The Collider Detector at Fermilab},"
North-Holland, 1988.}
and D0.\footnote{ D0 Collaboration, S. Abachi et al., to be published in Nucl.
Instrum. Methods A (1994).}  The
collaborations have reached a thorough understanding of their detectors,
and a detailed
correspondence of the observed signals for processes of interest with the
simulated
response from a model of the physics and the detector.  This enables both teams
to extrapolate the detector performance to higher energy and luminosity, and to
analyze the
upgrades which would be needed to reach specific signals for new phenomena.

        Both collaborations today are preparing Expressions of Interest for
a next
generation of physics in the detectors, beginning when the Injector Upgrade
will bring
about an order of magnitude increase in luminosity for the Tevatron (to
$10^{32}\,{\rm cm}^{-2}\,{\rm s}^{-1}$).
The likely conclusion of these EOI's is that both detectors would be fully
sensitive to the
above SUSY signals.  Operation at the higher luminosity will require faster
data
acquisition electronics and substantial upgrade of the tracking systems in CDF,
and the
addition of a magnet and tracking in D0.  These upgrades are already envisioned
and some
are in progress.  Otherwise both detectors are ready as they now operate to
seek SUSY at
the DiTevatron.

          The message is then clear.  {\bf Over the entire range of SUSY model
parameters, the DiTevatron could provide $\sim$80\% sensitivity for all of the
major sparticles in the SUSY spectrum.  The DiTevatron is for SUSY what the
SSC was to be for the Higgs.}
\vspace{.15in}

\noindent{\bf Uniqueness of discovery potential.}  A critical issue in
searching for the
signals for
supersymmetry is the struggle to optimize signal over background.  One
important source
of backgrounds which has already been encountered at CDF is the superposition
of several
distinct interactions in a single bunch-bunch crossing.  Confusion of
jets and leptons from
distinct interactions can produce spurious ``signals" for new physics
if many interactions
are superposed.  On average there are $\sim$2 interactions per
crossing in the Tevatron at its
current luminosity.  We have learned to cope with that level of
multi-interactions in the
searches for top and for SUSY.  As the Tevatron luminosity is
increased, the bunch
spacing will be reduced, so that at the ultimate luminosity of the
DiTevatron there would
still be only 3 interactions per bunch crossing.

        By contrast, LHC will operate with a luminosity of
$2.5\times10^{34}\,{\rm cm}^{-2}\,{\rm s}^{-1}$ and a
bunch spacing of $25\,{\rm ns}$, corresponding to {\bf 60} interactions per
bunch
crossing!  The push
for ultimate luminosity is essential if CERN is to maximize its reach
for the Higgs boson.
In the ``golden" signal for the Higgs (4 leptons from 2 $Z$ decays), the
additional pair-mass
constraints may make it feasible to reject combinatoric backgrounds.  For
supersymmetry,
however, we have no such constraints.  Hence the irony: although LHC will
have ample
energy and luminosity to make plenty of SUSY particles, it
could be
extremely difficult to find them.  The DiTevatron is a superior machine for
the purpose.
The collider and its detectors could be ready to operate for physics in five
years.
\vspace{.15in}

\noindent{\bf The DiTevatron would also be a top factory.}
Anticipating that the top quark will be
discovered during the next run at Fermilab, the production cross-section
would be 20
times larger at twice the collision energy.  Taking this increase together
with the 20-fold
increase of luminosity afforded by the Injector Upgrade, the CDF and D0
experiments at
the DiTevatron would each be able to identify and reconstruct $\sim$4,000
top quarks/year.
Such a data set would make it possible to make detailed studies of the
last quark: its
mass, its decay modes, the electroweak interaction with a quark which is
heavier than the
gauge bosons themselves.
\vspace{.15in}

\noindent{\bf A World Vision.}
We conceive a world vision for cooperation in high energy physics,
which recognizes the legitimate aspirations of each continent's high energy
community, while working together to best use of resources.

        The DiTevatron has a unique potential for a milestone discovery of
historic
significance - supersymmetry.  It could be built in five years, as a final,
constructive
chapter in the termination of the SSC.  Its capital cost has been roughly
estimated by
Fermilab at $\sim$\$250 million.  In the family of \$billion facilities planned
in Europe and Japan,
it could prove to be an excellent bargain for Yankee ingenuity, in the same
way that the
Tevatron itself bootstrapped a $400\,{\rm GeV}$ accelerator into a $2\,{\rm
TeV}$ collider
which is today
the sole source of top quarks.

        Europe plans to build LHC.  It has excellent discovery potential for
the Higgs
boson, and could be complete in a decade.  There should be opportunity for
collaboration
by universities and laboratories world-wide in the technology development and
physics
program of the LHC, driven by the project's needs and the scientists'
interests.

        Both Japan and SLAC plan to build $e^+e^-$ $B$ factories.
These facilities will
complement the DiTevatron to measure the parameters of CP violation.  Together
with the $\varphi$
factory at INFN Frascati, the results may lead us to an understanding
of the origins of CP violation.

        The fixed-target programs at Fermilab and CERN have for 25 years
been a stalwart component of the HEP scene.  With the advent of the Tevatron
upgrade, new
experiments with rare $K$ decays and long-baseline neutrino oscillations
will dramatically
extend earlier studies.  Effects from all three of the above
milestone-discovery interactions
could be tested in these two experiments.

        Japan has just announced plans to build an $e^+e^-$ linac collider,
beginning in 1996.
The first stage, $150\,{\rm GeV}$/beam, could well prove to be an excellent
facility
for producing
top quarks as well as probing for an intermediate mass range of the Higgs
boson.  The
ultimate stage, $\sim500\,{\rm GeV}/$beam, would support a detailed study of
SUSY
Higgs physics,
and of the superpartners of the leptons.  The latest suggestions of converting
the particle
beams into photon beams just before collision may pose an even more elegant
physics
channel.  In any case, the linac collider will be a {\it tour de force} of
accelerator technology,
requiring a sustained international effort to prepare for it.  There are
today a number of
international teams working to mature its technology: SLAC, CLIC, TESLA,
and INP
Novosibirsk.  It is to be hoped that KEK will pull together an effective
teamwork of these
talents to mature the best possible design and its component technologies.
\vspace{.15in}

\noindent{\bf Technology - the art of the possible.}  As we begin to build
accelerators and detectors for
the next generation of high energy physics experiments, we should realize that
we are relying on the
fruits of a long, sustained effort of technology development.  The innovations
of beam
cooling, high-$j_c$ superconducting strand, 8 Tesla dipoles, nano-focus beams,
suppression
of wake fields, high-power microwave drivers, AC-coupled silicon microstrip
and pixel
detectors, ring-imaging Cerenkov counters, scintillating fibers, and gas
microstrip
chambers have required years of effort and patient support.  As we begin
to use these
technologies, it is equally important that we do a more effective job of
supporting the
further innovation of new technologies that can provide an equally rich armada
when it
comes time a decade hence to ask again - what should we do next?  At each such
juncture,
our vision is bounded by what we know to be possible; technology is the means
of pushing back those bounds.

        One example is of particular relevance to future hadron colliders.
Two years ago,
one of us (PM) co-invented a wholly new approach to high-field dipole design
- the pipe magnet\footnote{ E. Badea, P.M. McIntyre, and S. Pissanetzky, ``The
Pipe Magnet: Compact 13 Tesla Dual Dipole for Future Hadron Colliders," Proc.
HEACC '92 Conference, Hamburg (1992).} shown in Figure~\ref{Figure8}.
It is a 13 Tesla dual dipole, occupying the same space as one
SSC dipole.  The magnet is configured essentially as a deformed toroid with
two dipole
insertions.  We used a conformal mapping to design a flux transformer which
produces collider-quality field in each insertion.  The magnetic flux is guided
everywhere around its circuit, reducing
peak stress in the coil by a factor 2 and eliminating persistent-current
multipoles - two of
the most serious limits to $\cos\theta$ magnet performance.
The pipe magnet also incorporates
innovations in ${\rm Nb}_3{\rm Sn}$ coil fabrication, prestress distribution,
and cooling
which we have
learned from experience with other field geometries.

        The best experts in magnet design and fabrication at Fermilab and CERN
have
expressed enthusiasm for the concept.  Nevertheless it is only a concept.  It
would take
$\sim$5 years of intense effort by a seasoned team to build and test model
magnets, and
strive to mature it to readiness as a collider magnet.  We have such a team
and six months
ago we requested funding from DOE's Advanced Accelerator R\&D Program for this
purpose.  The answer: there is no money in all of DOE HEP for any new programs.

        If we take such an attitude to the first new technology for
superconducting dipoles
in twenty years, and to similar new technologies for accelerators and
detectors, we will have no future as a field!
\vspace{.15in}

\noindent{\bf The role of the universities.}
As the accelerators and detectors required for
our research
grow ever larger and more complex, there is a natural tendency to consolidate
the design,
construction, operation, analysis, and supporting technology R\&D at major
national
laboratories, where the tools supporting technical personnel and facilities
can assure a
competent, well-managed effort.

        Beware!  Look back at Table I, and consider the people who made
those milestone
discoveries.  With very few exceptions, they were faculty at universities.
The universities
of America are one of our greatest assets as a nation.  The high energy
physicists on their
faculties are the driving force of creativity and sustained effort for
the research which our
high energy facilities are being conceived to support.  They are also
the originators of
most of the new concepts for accelerator and detector technologies that
extend the
possible.  Yet over the past decade there has been a steady erosion of the
support of these
scientists within the U.S. high energy program.  If we continue such a shift
in our
priorities, we will have no future as a field!
\vspace{.15in}

\noindent{\bf The future is bright.}  The SSC is dead.  We grieve for its
passing.  We
hope to have
demonstrated a vision for high energy physics in which the future can be
bright.  We are at
the threshold of a milestone discovery (the top quark), and we have the
realizable potential
within our austere budget to reach for another one - supersymmetry.  The
DiTevatron
could be operating for research within five years.  Its construction is
compatible with the
current effort to discover the top quark - the next milestone discovery.  It
fits well into a
world view in which Europe builds an LHC, Japan builds a linac collider, and
international
collaboration flourishes for technology development and physics experiments.
We could
accomplish this agenda within our likely budget constraints, but only if we
develop a
better ability to balance our priorities.  The waning support of technology
R\&D and of our
universities would jeopardize the field's future were it to continue.
\newpage

{
\oddsidemargin -0.4in
\evensidemargin 0in
\textwidth 6.9in
\textheight 9.0in
\topmargin -0.7in
\parindent 0.5in

\noindent {\large\bf Table 1: Highlights in Particle Physics}
$$\!\!\!\!\!\!\!\!\!\!\!\!\!\!
\begin{array}{llll}
 \mbox{Year} & \mbox{Discovery} & \mbox{Source} &
				\mbox{Anticipated Theory}  \\
\hline \hline
 \mbox{1896} & \alpha, \beta, \gamma & \mbox{Radioactive Source} &  \\ \hline
 \mbox{1897} & e^{-}, p  & \mbox{Cathode Ray Tube, H$^{+}$ }  &     \\ \hline
 \mbox{1911} & \mbox{Rutherford scattering}
			& \mbox{$\alpha$ + Au (gold foil)} &
			\mbox{Rutherford's atom} \\ \hline
 \mbox{1932} & n    &
	\mbox{$\alpha$ + $^{9}$Be$_{4}$ $\rightarrow$ $n$ + $^{12}$C$_{6}$} &
								    \\ \hline
 \mbox{1933} & e^{+} 	 	& \mbox{Cosmic Ray} &
					\mbox{Dirac Theory (1928)}  \\ \hline
 \mbox{1937} & \mu^{\pm} 	& \mbox{Cosmic Ray} &		    \\ \hline
 \mbox{1947} & \pi^{\pm} 	& \mbox{Cosmic Ray} &
					\mbox{Yukawa Theory (1935)} \\ \hline
 \mbox{1947} & K^{\pm,0} 	& \mbox{Cosmic Ray} &		    \\ \hline
 \mbox{1950} & \pi^{0} 	& \mbox{Cosmic Ray}	&	\\
	     &		& \mbox{SC (184-in., LBL)} & \\ \hline
 \mbox{$\sim$1953} & \mbox{Hyperons} & \mbox{Cosmic Ray} & \mbox{Strangeness}
\\
	     & \Lambda^{0}, \Sigma^{+}, \Xi^{-} &
	\mbox{1.4-GeV/c $\pi^{-}$ by PS (BNL)} &
		\mbox{Gell-Mann (1953), Nishijima (1955)} \\
 \mbox{(1959)} & (\Sigma^{0}, \Xi^{0}) &  	& 	\\ \hline
 \mbox{1955} & \overline{p} & \mbox{Bevatron (6.3 GeV/c $p$, LBL)} &
				\mbox{Dirac Theory (1928)} \\ \hline
 \mbox{1959} & \nu_{e} & \mbox{Reactor ($\beta$ decay)} &
		 	\mbox{Pauli's invention of $\nu$ (1930)} \\
             &   &   & 	\mbox{Fermi's $\beta$-decay Theoty (1934)} \\ \hline
 \mbox{1962} & \nu_{e} \neq \nu_{\mu} & \mbox{AGS (BNL)} &
		\mbox{No events for
		$\mu^{-} \rightarrow e^{-} \; \nu \; \overline{\nu}
		\rightarrow e^{-} \; \gamma$ } \\ \hline
 \mbox{1964} & \Omega^{-} & \mbox{AGS (BNL)} &
				\mbox{SU(3) - Eightfold Way} \\
             &       &       &	\mbox{Gell-Mann (1961), Ne'eman (1961)} \\
             &       &       &  \mbox{Quark Model ($u,d,s$)} \\
             &       &       &	\mbox{Gell-Mann (1964), Zweig (1964)} \\ \hline
 \mbox{1973} & \mbox{Neutral Current} & \mbox{SPS (400 GeV $p$, CERN)}&
			     SU(2) \times U(1) \\
             & & & \mbox{Glashow (1961)}	\\
             & & & \mbox{Weinberg (1967), Salam (1968)} \\ \hline
 \mbox{1974} & J/ \psi & \mbox{SPEAR ($e^{+}e^{-}$, SLAC)}&
						\mbox{Charm Quark} \\
	     &         & \mbox{AGS (28 GeV $p$, BNL)} &
			\mbox{Glashow, Iliopoulos, Maiani (1970)} \\ \hline
 \mbox{1975} & \tau  & \mbox{SPEAR ($e^{+}e^{-}$, SLAC)} &
					R_{exp} - R_{theory} \sim 1;
		R = \sigma_{h} / \sigma_{\mu \mu}  \\ \hline
 \mbox{1977} & \Upsilon & \mbox{PS (400 GeV $p$, FNAL)}&
		\mbox{Kobayashi-Maskawa Theory (1972)} \\
	     &          &      &  \mbox{(CP Violation)}	\\ \hline
 \mbox{1979} & g & \mbox{PETRA (31 GeV $e^{+}e^{-}$, DESY)}&
			\mbox{QCD ($SU(3)_{c}$)}\\ \hline
 \mbox{1983} & W^{\pm}, Z^{0} &
		\mbox{S$p\overline{p}$S (540 GeV $\overline{p}p$, CERN)}&
				SU(2) \times U(1) \\ \hline
\mbox{1989} & N_g=3 & \mbox{LEP (91 GeV $e^+e^-$, CERN)} & \mbox{GUTs}\\ \hline
 \mbox{1994?} & top & \mbox{Tevatron (1.8 TeV $\overline{p}p$, FNAL) } &
			\mbox{Kobayashi-Maskawa Theory (1972)} \\
         &  & & \mbox{No anomalies in $SU(2) \times U(1)$} \\
         &  & & \mbox{Suppression of FCNC} \\ \hline
 \mbox{1999?} & SUSY & \mbox{SUSYtron (4 TeV $p\bar{p}$, FNAL) ?} &
				\mbox{SUSY + GUT} \\ \hline
 \mbox{2005?} & Higgs & \mbox{LHC (14 TeV $pp$, CERN) } &
				\mbox{Standard Model} \\
	     & \mbox{{\it New Particles}} & 	&
				\mbox{Beyond the Standard Model} \\ \hline
\end{array}
$$
\begin{flushleft}
Notes: SC = Synchrocyclotron,  PS = Proton Synchrotron
\end{flushleft}
}
\newpage

\begin{figure}[p]
\vspace{6.0in}
\caption{Cross-section of the Fermilab tunnel, showing the DiTevatron (bottom),
the Tevatron (middle), and the transport line (top) in A sector for transfer
of beams from the 150 GeV Injector. The beam height for collisions is the
same as for the present Tevatron. }
\label{Figure1}
\end{figure}

\begin{figure}[p]
\vspace{7.0in}
\includegraphics{Fig2a.ps}
\vspace{-0.5in}
\caption{End projection of a trilepton event in the CDF experiment. The event
contains two CMU+CMP muon candidates, plus one CMI candidate. The muons are
visible as straight tracks on the right ($\mu^+$, $p_\perp=35\,{\rm GeV}$),
the lower left ($\mu^-$, $p_\perp=10.5\,{\rm GeV}$), and a short track in the
upper left which exits into the end plug ($\mu^-$, $p_\perp=49\,{\rm GeV}$
imperfectly measured).}
\label{Figure2a}
\end{figure}

\begin{figure}[p]
\vspace{3in}
\includegraphics{Fig2bI.ps}
\vspace{4in}
\includegraphics{Fig2bII.ps}
\vspace{0.75in}
\caption{Side views of $ee\mu$ trilepton candidate from the D0 experiment.
The event contains a muon at $\eta$ = 1.1 and measured $p_\perp$ = 17.2
GeV/c and two electron candidates (seen in the end cap)  with $\eta$'s of 2.0
and 1.9 and $p_\perp$'s of 37.8 and 8.0 GeV/c respectively. The missing
$p_\perp$ is 39.1$\pm$ 2.7 GeV/c without, and 23.3 $\pm$ 11.6 GeV/c with the
muon included.}
\label{Figure2b}
\end{figure}

\begin{figure}[p]
\vspace{3.5in}
\includegraphics{Fig3.ps}
\vspace{1.5in}
\caption{Trilepton yield ($\sigma B$) versus chargino mass in chargino
production. The dots define the range of parameters allowed within the
standard SU(5) supergravity model. Results are shown for each sign of the
Higgs mixing parameter $\mu$. The upper plots show the limits which could
be reached in the Injector-upgraded Tevatron; the lower plots show the
limits which could be reached in the DiTevatron.}
\label{Figure3}
\end{figure}

\begin{figure}[p]
\vspace{3.5in}
\includegraphics{Fig4.ps}
\vspace{1.5in}
\caption{Trilepton yield ($\sigma B$) versus chargino mass in chargino
production. The dots define the range of parameters allowed within a no-scale
SU(5)xU(1) supergravity model using moduli supersymmetry-breaking at the GUT
scale. Results are shown for each sign of the Higgs mixing parameter $\mu$. The
upper plots show the limits which could be reached in the Injector-upgraded
Tevatron; the lower plots show the limits which could be reached in the
DiTevatron.}
\label{Figure4}
\end{figure}

\begin{figure}[p]
\vspace{3.5in}
\includegraphics{Fig5.ps}
\vspace{1.5in}
\caption{Trilepton yield ($\sigma B$) versus chargino mass in chargino
production. The dots define the range of parameters allowed within a no-scale
SU(5)xU(1) supergravity model using dilaton supersymmetry-breaking at the GUT
scale. Results are shown for each sign of the Higgs mixing parameter $\mu$. The
upper plots show the limits which could be reached in the Injector-upgraded
Tevatron; the lower plots show the limits which could be reached in the
DiTevatron.}
\label{Figure5}
\end{figure}

\begin{figure}[p]
\vspace{7.0in}
\includegraphics{Fig6.ps}
\caption{Spectrum of missing $p_\perp$ for events containing 4 jets (each with
$p_\perp>40\,{\rm GeV}$). Spectra are shown for signal events from 400 GeV
gluinos (dashed curve), 500 GeV gluinos (dotted curve), and background
from $Z\to \nu\bar\nu$.}
\label{Figure6}
\end{figure}

\begin{figure}[p]
\vspace{5.0in}
\includegraphics{Fig7.ps}
\vspace{1.7in}
\caption{Statistical significance for gluino and squark events
selected by the criteria $p_\perp>150\,{\rm GeV}$, and 4 jets with
$p_\perp>40\,{\rm GeV}$. Bands are shown for signal $S$ from gluino pairs
and squark/gluino combinations for parameters which are consistent with the
standard SU(5) supergravity model and the SU(5)xU(1) supergravity
models, respectively. The background $B$ is calculated from $Z\to\nu\bar\nu$;
the bands provide for a factor of 5 deterioration of $S/B$ ratio
due to additional backgrounds or inefficiencies.}
\label{Figure7}
\end{figure}

\begin{figure}[p]
\vspace{7.0in}
\caption{Configuration of the pipe magnet in a cryostat appropriate for
hadron collider applications. The magnet would produce a 13 Tesla dipole
field in each beam tube, with multipoles suitable for collider operation.
An intercept tube for synchrotron light is shown.}
\label{Figure8}
\end{figure}

\end{document}